\documentclass{ws-p8-50x6-00}

\newcommand{\bda}{\begin{\displaymath}\begin{array}{rl}}
\newcommand{\eda}{\end{array}\end{displaymath}}
\newcommand{\bdm}{\begin{displaymath}}
\newcommand{\edm}{\end{displaymath}}
\newcommand{\no}{\nonumber \\}
\newcommand{\fs}{\; \; .}
\newcommand{\co}{\; \; ,}
\newcommand{\all}{&\!}

\newcommand{\lbar}{\bar{\ell}}

\newcommand{\Wbar}{\,\overline{\rule[0.75em]{0.9em}{0em}}\hspace{-1em}W}
\newcommand{\Pbar}{\,\overline{\rule[0.75em]{0.5em}{0em}}\hspace{-0.7em}P}

\newcommand{\pbar}{\overline{\rule[0.5em]{0.4em}{0em}}\hspace{-0.5em}p}

\newcommand{\Ezero}{\sqrt{\rule[0.1em]{0em}{0.5em}s_0}}

\newcommand{\rs}{\langle r^2\rangle\rule[-0.2em]{0em}{0em}_s}

\newcommand{\gev}{\mbox{GeV}}
\newcommand{\mev}{\mbox{MeV}}

\begin{document}

\title{Theory of $\pi \pi$ scattering}

\author{Gilberto Colangelo}

\address{Institut f\"ur Theoretische Physik der Universit\"at Z\"urich,\\
Winterthurerstr. 190\\
CH--8057 Z\"urich}

%%%%%%%%%%%%%%%%%%%%%%%%%%%%%%%%%%%%%%%%%%%%%%%%%%%%%%%%%%%%%%
% You may repeat \author \address as often as necessary      %
%%%%%%%%%%%%%%%%%%%%%%%%%%%%%%%%%%%%%%%%%%%%%%%%%%%%%%%%%%%%%%

\maketitle

\abstracts{I describe the current status of the theory of $\pi \pi$
  scattering, reviewing in particular recent work on the numerical solution
  of Roy equations and on the matching between these and the chiral
  representation. I discuss numerical results on the scattering lengths and
  other threshold parameters.} 

\section{Introduction}
The study of $\pi \pi$ scattering is a classical subject in the field of
strong interactions: it is a scattering process that may occur at very low
energy, far away from the resonance region, and it only involves the
quasi--Goldstone bosons of the chiral symmetry of strong interactions. If we
stay at low energy, we could expect its dynamics to be strongly constrained
by chiral symmetry: as Weinberg showed 
\cite{weinberg66}, in the chiral limit, the scattering amplitude has to
vanish when the momenta of the pions tend to zero.  In the real world
quarks are not massless, and consequently also the would--be Goldstone
bosons acquire a small mass. The chiral argument given above implies that
the amplitude at threshold must be proportional to the square of the mass
of the pions \cite{weinberg66}: \be
\label{eq:weinberg}
a_0^0=\frac{7 M_\pi^2}{32 \pi F_\pi^2}, \; \; a_0^2=-\frac{M_\pi^2}{16 \pi
F_\pi^2} \; , 
\ee 
where $a_\ell^I$ stands for the scattering length in
the isospin $I$ channel with angular momentum $\ell$.

The formulae above are valid only at leading order in a series expansion in
powers of the quark masses: the next--to--leading order corrections have
been calculated by Gasser and Leutwyler \cite{GL 1983}, and even the
next--to--next--to--leading order corrections are now known \cite{BCEGS}. 
The latter had been calculated already before the previous edition of the
``Chiral Dynamics'' series of Workshops, and the interested reader may find
a report of the situation at that time in the contribution by Gerhard Ecker
to the previous Workshop Proceedings \cite{ecker CD97}.
The main difference between that report and the present one is the fact
that at that time the numerical analysis had not been completed: in order to
illustrate the numerical size of the two--loop contributions, Ecker
had to present three different numerical values of the $S$--wave scattering
lengths, depending on the input used for the low energy constants (see
table 1 in Ref. \cite{ecker CD97}). The numerical analysis has been
recently completed \cite{CGL}, and has yielded the values:
\be
\label{eq:final numbers}
a_0^0=0.222 \pm 0.005 , \; \; a_0^2=-0.0445 \pm 0.001 \; .
\ee
In what follows I will describe what are the experimental ingredients and the
theoretical tools that have lead to this result.

\section{Numerical solutions of Roy equations}
In 1971 Roy \cite{roy} derived a set of dispersion relations for the
partial wave amplitudes of $\pi \pi$ scattering, that fully respected
crossing symmetry, and that involved only two
subtraction constants. The crucial observation that lead to this result was
that if one writes a fixed--$t$ dispersion relation, crossing symmetry
constrains the form of the $t$ dependent subtraction constants in such a
way that they can be expressed in terms of the two $S$--wave scattering
lengths and other dispersive integrals. The equations have the following
form:
\be\label{eq:req1} t_\ell^I(s)= k_\ell^I(s)+
\sum_{I'=0}^2\sum_{\ell'=0}^\infty \int_{4M_\pi^2}^\infty
ds'\,K_{\ell\ell'}^{I I'}(s,s')\,\mbox{Im} \, t_{\ell'}^{I'}(s')\, ,
\ee
where $I$ and $\ell$ denote isospin and angular
momentum, respectively and $k^I_\ell(s)$ is the partial wave 
projection of the subtraction term, present only in $S$- and $P$-waves,
\bea 
\label{eq:subconst1}
\hspace{-6mm}
k^I_\ell(s)\all=\all a_0^I \,\delta_\ell^0 +\frac{s-4M_\pi^2}{4M_\pi^2}\,
(2a_0^0-5a_0^2)\,\left(\frac{1}{3}\,\delta_0^I\,\delta_\ell^0
+\frac{1}{18}\,\delta_1^I\,
\delta_\ell^1-\frac{1}{6}\,\delta_2^I\,\delta^0_\ell\right).
\eea
The kernels $K_{\ell\ell'}^{I I'}(s,s')$ are explicitly known
functions. They contain a diagonal, singular Cauchy kernel that 
generates the right hand cut in the partial wave amplitudes, as well as a
logarithmically singular piece that accounts for the left hand
cut.
Soon after these equations became available in the literature, various groups
started to treat them numerically \cite{roy-num}.
At that time the goal was to derive from the available data constraints on
the scattering lengths: all groups agreed on the result that only data
sufficiently close to threshold could provide significant bounds on the
scattering lengths. With the only exception of data on $K_{e4}$ decays, all
other data were at too high an energy, and therefore left the $S$--wave
scattering lengths practically unconstrained.

The only high--statistics experiment on $K_{e4}$ decays provided its final
results in 1977 \cite{rosselet}, and it indeed allowed (when combined with
the numerical studies of Roy equations) to constrain the value of the $I=0$
$S$--wave scattering length to within a rather narrow range:
$a_0^0=0.26\pm0.05$. Until today this was the best experimental information
available on this observable. Now we have already preliminary results
\cite{e865} from a new experiment that has been operating at Brookhaven,
analysing more than $4\cdot 10^5$ $K_{e4}$ decays (to be compared to the
$3\cdot 10^4$ of the previous experiment \cite{rosselet}).

The prospects of a new generation of high--statistics experiment on
$K_{e4}$ decays and theoretical work on the calculation of the scattering
lengths made necessary a revival of the numerical treatment of Roy
equations. An extensive work has been made in \cite{ACGL}, whose highlights
I will now briefly summarize:

\noindent
1. Given the strong dominance of the $S$-- and $P$--waves at low energy,
Roy equations have been solved only for these, and only on the interval
$4M_\pi^2<s<s_0=(0.8 \; \gev)^2$, the lower half of their range of validity
(which has been rigorously proven to extend up to $\sqrt{s_1}=1.15
\; \gev$). In that region the contributions generated by inelastic channels
can be safely neglected. In the interval from $s_0$ to $s_2=(2 \; \gev)^2$,
the imaginary parts have been evaluated with the available experimental
information, whereas above $s_2$, a theoretical representation, based on
Regge asymptotics has been used.

\noindent
2. Unitarity converts the Roy equations for the $S$- and $P$-waves into a
set of three coupled integral equations for the corresponding phase shifts:
The real part of the partial wave amplitudes is given by a sum of known
contributions (subtraction polynomial, integrals over the region
$s_0<s<s_2$ and driving terms) and certain integrals over their imaginary
parts, extending from threshold to $s_0$. Since unitarity relates the real
and imaginary parts in a nonlinear manner, these equations are inherently
nonlinear and cannot be solved explicitly.

\noindent
3. Several mathematical properties of such integral equations are known: In
particular, the existence and uniqueness of the solution is guaranteed only
if the matching point $s_0$ is taken in the region between the place where
the $P$-wave phase shift goes through $90^\circ$ and the energy where the
$I=0$ $S$-wave does the same. As this range is quite narrow
($0.78\,\mbox{GeV}<\Ezero< 0.86\,\mbox{GeV}$), there is little freedom in
the choice of the matching point.

\noindent
4. A second consequence of the mathematical structure of the Roy equations
is that, for a given input and for a random choice of the two subtraction
constants, the solution has a cusp at $s_0$: In the vicinity of the
matching point, the solution in general exhibits unphysical behaviour. The
strength of the cusp is very sensitive to the value of $a^2_0$. In fact, it
was found that the cusp disappears if that value is tuned
properly. Treating the imaginary parts as known, the requirement that the
solution is free of cusps at the matching point determines the value of
$a_0^2$ as a function of $a_0^0$.

\noindent
5. The input used for the imaginary parts above the matching point is
subject to considerable uncertainties. In this framework, the values of the
$S$- and $P$-wave phase shifts at the matching point represent the
essential parameters in this regard. The data on the pion form factor,
obtained from the processes $e^+e^-\rightarrow \pi^+\pi^-$ and
$\tau\rightarrow \pi^-\pi^0\nu_\tau$, very accurately determine the
behaviour of the $P$--wave phase shift in the region of the
$\rho\,$--resonance, thus constraining the value of $\delta_1^1(s_0)$ to a
remarkably narrow range. The phase shifts extracted from the reaction $\pi
N\rightarrow\pi\pi N$ constrain rather strictly the difference
$\delta_0^0-\delta_1^1$, better than either of the two phases individually.
Since the $P$--wave is known very accurately from the leptonic processes
mentioned above, this implies that $\delta_0^0(s_0)$ is also known rather
well. The experimental information concerning $\delta_0^2$, on the other
hand, is comparatively meagre.

\noindent
6. The uncertainties in the experimental input for the imaginary parts and
those in the driving terms turn the universal curve into a band in the
$(a_0^0,a_0^2)$ plane. Outside this ``universal band'', the Roy equations
do not admit physically acceptable solutions that are consistent with what
is known about the behaviour of the imaginary parts above the matching
point.

\noindent
7. The Olsson sum rule relates the combination $2a_0^0-5a_0^2$ of
scattering lengths to an integral over the imaginary parts of the
amplitude. Evaluating the integral, it was found that the sum rule is
satisfied within a band that has a large overlap with the ``universal
band'' mentioned above. It is by no means built in from the start that the
two requirements can simultaneously be met -- the fact that this is the
case represents a rather thorough check of our analysis.

\noindent
8. The admissible region can be constrained further if use is made of
experimental data below the matching point.  At the moment there are two
main sources of information on $\pi \pi$ scattering below 0.8 GeV: A few
data points for the $I=2$ $S$-wave phase shift -- which will,
unfortunately, not be improved in the foreseeable future -- and a few data
points on $\delta_0^0-\delta_1^1$ very close to threshold, from
$K_{e4}$ decays. The latter provide an important constraint. 

\noindent
9. The Roy equation analysis is the only method that allows one to reliably
translate low--energy data on the scattering amplitude into values for the
scattering lengths. As discussed above, the available data do correlate the
value of $a^2_0$ with the one of $a^0_0$.  Unfortunately, however, the
value of $a_0^0$ is not strongly constrained: In agreement with
earlier analyses, it was found that these data are consistent with any
value of $a_0^0$ in the range from 0.18 to 0.3.

\noindent
10. The two subtraction constants $a_0^0$, $a_0^2$ are the essential
parameters at low energies: If these were known, the method would allow one
to calculate the $S$- and $P$-wave phase shifts below 0.8 GeV to an amazing
degree of accuracy. In particular one can evaluate the amplitude also at
unphysical points. This is very important if one wants to use at best the
information contained in the chiral representation of the $\pi \pi$
scattering amplitude, as we will see in the following section.

\section{Matching the chiral and the dispersive representation}
The Weinberg formula for the $S$--wave, $I=0$ scattering lengths
(\ref{eq:weinberg}) is subject to substantial corrections. The physical
reason for the size of these corrections is the strong rescattering of
pions in the $I=0$ channel. The leading order formula fails to fully
describe such effects because unitarity corrections only show up at
next--to--leading order and beyond. On the other hand, if one uses the
Weinberg amplitude below threshold and away from it, these unitarity
effects will become less and less important, and the nominal parameter of
the chiral expansion ($M_\pi^2$ in 1 GeV$^2$ units, a number of the order
of a few percent) will start to dictate the actual size of the corrections.

The fact that in the Roy equations the $S$--wave scattering lengths appear
as the subtraction constants is only a matter of choice: one could as well
subtract at an unphysical point, without changing any of the physical
results. A combination of the two approaches that exploits the strengths of
both is therefore possible: the chiral expansion may provide accurate
information on the two subraction constants, by fixing the value of the
amplitude way below threshold, but would be less precise in the evolution
of the amplitude up to threshold and above -- there the dispersive
framework based on the solution of Roy equations is much better.  This
program has been carried out in Ref. \cite{CGL}, making full use of the
information contained in the two--loop chiral representation of the
amplitude. In the following I will briefly describe some of the details.

\subsection{Chiral representation of the scattering amplitude}
The two-loop representation yields the first three terms in the low energy
expansion of the partial waves:
\be\label{seriespw} t^I_\ell(s)=t^I_\ell(s)_2+ t^I_\ell(s)_4+
t^I_\ell(s)_6+O(p^8)\fs \ee
Since inelastic reactions start showing up only at
$O(p^8)$, unitarity implies
\bea \mbox{Im}\,t^I_\ell(s)=\sigma(s)\,|\hspace{0.04em}t^{I}_\ell(s)
\hspace{0.03em}|^2+O(p^8)\co\hspace{2em}
\sigma(s)= (1-4M_\pi^2/s)^{\frac{1}{2}}
\fs\eea
The condition fixes the imaginary parts of the two-loop
amplitude in terms of the one-loop representation. At leading order, 
the scattering amplitude is linear in the Mandelstam variables, so that
only the S- and P-waves are different from zero. Unitarity therefore
implies that, up to and including $O(p^6)$, only these partial waves
develop an imaginary part. Accordingly, the chiral representation 
of the scattering amplitude can be written as~\cite{Knecht Moussallam
  Stern Fuchs}   
\bea\label{chiral decomposition} A(s,t,u)
\all=\all C(s,t,u)+32\pi\left\{\mbox{$\frac{1}{3}$}\,U^0(s)+
\mbox{$\frac{3}{2}$}\,(s-u)\,U^1(t)
+\mbox{$\frac{3}{2}$}\,(s-t)\,U^1(u)\right.\no
\all\all \left.+\mbox{$\frac{1}{2}$}\,U^2(t)+\mbox{$\frac{1}{2}$}\,U^2(u)
-\mbox{$\frac{1}{3}$}\, U^2(s) \right\}+ O(p^8)\co\eea
where $C(s,t,u)$ is a crossing symmetric polynomial,
\be\label{poly}
C(s,t,u)=c_1+s\,c_2+s^2\,c_3
+(t-u)^2\,c_4+s^3\,c_5
+s\,(t-u)^2\,c_6\fs \ee
The functions $U^0(s)$, $U^1(s)$ and $U^2(s)$ describe the 
``unitarity corrections'' associated with $s$-channel isospin $I=0,1,2$,
respectively. In view of $\mbox{Im}\,t^I_\ell(s)_6\propto s^3$, several
subtractions are needed for the dispersive representation of these functions
to converge. We subtract at $s=0$ and set
\bea\label{disp U}U^I(s)\all=\all \frac{s^{4-\epsilon_I}}{\pi}
\int_{4M_\pi^2}^\infty ds'\;\frac{\sigma(s')\,t^I(s')_2\,\{t^I(s')_2+ 2\,
\mbox{Re} \,t^I(s')_4\} }
{s^{\prime\,4-\epsilon_I}(s'-4M_\pi^2)^{\epsilon_I}(s'-s )}\fs\eea
The subtraction constants are collected in $C(s,t,u)$.
As only the $S$- and $P$-waves enter, we have dropped the lower index,
$\{t^0,t^1,t^2\}=\{t^0_0,t^1_1,t^2_0\}$. For kinematic reasons, the
integrand of the $P$-wave differs from the one of the $S$-waves:
$\{\epsilon_0,\epsilon_1,\epsilon_2\}=\{0,1,0\}$.
It is straightforward to check that the result of the two-loop 
calculation \cite{BCEGS} is indeed of this structure. 

The two--loop result for the amplitude specifies how the coefficients
$c_1,\ldots,\,c_6$ depend on the quark masses~\cite{BCEGS} (for explicit
expressions, see \cite{CGL2}). These formulae, in particular contain
Weinberg's low energy theorems, which in this language state that the
leading terms in the expansion of the first two coefficients are fixed by
$M_\pi$ and $F_\pi$: 
$c_1=-M_\pi^2/F_\pi^2+\ldots\,$, $c_2=1/ F_\pi^2+\ldots$ At first order,
the constants $\ell_1,\ell_2,\ell_3,\ell_4$ from ${\cal L}_4$ \cite{GL 1984}
enter, and at second order, the chiral representation of the 
scattering amplitude involves the couplings $r_1,\ldots\,,r_6$ from ${\cal
L}_6$ \cite{BCE}.  It is useful to distinguish two categories of coupling
constants:
\begin{description}\item {\it a. Terms that survive in
the chiral limit.} Four of the coupling constants that enter the two-loop
representation of the scattering amplitude belong to this category:
$\ell_1,\,\ell_2,\,r_5,\,r_6$.
\item {\it b. Symmetry breaking terms.}
The corresponding vertices are proportional to a power
of the quark mass and involve the coupling constants $\ell_3$, $\ell_4$,
$r_1$, $r_2$, $r_3$, $r_4$.\end{description}
The constants of the first category show up in the momentum dependence of 
the scattering amplitude, so that these couplings may be determined
phenomenologically. The symmetry breaking terms, on the other hand,
specify the dependence of the amplitude on the quark masses. Since these 
cannot be varied experimentally, information concerning the second category
of coupling constants can only be obtained from sources other than $\pi\pi$
scattering. The constants $r_n$ from ${\cal L}_6$
only generate tiny effects, so that crude theoretical estimates
suffice, but the couplings $\ell_3$ and $\ell_4$ from ${\cal L}_4$
do play an important role.

The crucial parameter that distinguishes the standard framework from the
one proposed in Ref.~\cite{Stern Sazdjian Fuchs} is $\ell_3$. This coupling
constant determines the first order correction in the
Gell-Mann-Oakes-Renner-relation: $ M_\pi^2= 2 B
m\left\{1-\frac{1}{2}\xi\lbar_3 +O(\xi^2)\right\}$, $\xi=2Bm/(16 \pi^2
F_\pi^2)$.  The value of $\ell_3$ is not known accurately.  Numerically,
however, a significant change in the prediction for the scattering lengths
can only arise if the crude estimate in Ref.~\cite{GL 1984},
\be\label{l3barnum} \bar{\ell}_3=2.9\pm 2.4\,,\ee should turn out to be
entirely wrong. We do not make an attempt at reducing the uncertainty in
$\ell_3$ within the standard framework, but will explicitly indicate the
sensitivity to this coupling constant.

Chiral symmetry implies that the coupling constant $\ell_4$ also shows up in 
the expansion of the scalar radius in powers of the quark
masses \cite{GL 1983}: 
\bea \rs = \frac{3}{8\pi^2 F_\pi^2}
\left\{\lbar_4-\frac{13}{12}+\xi\,\Delta_r+O(\xi^2)\right\}\fs
\eea
As pointed out in Ref.~\cite{DGL}, the scalar radius can be determined
on the basis of a dispersive evaluation of the scalar form factor.
The result~\cite{CGL}
\be\label{rsnum} \rs=0.61\pm 0.04\,\mbox{fm}^2\co\ee 
is an update of the value given in Ref.~\cite{DGL} and is consistent with
earlier  estimates of the low energy constant $\ell_4$ based on the
symmetry breaking seen in $F_K/F_\pi$ or on the decay $K\rightarrow
\pi\ell\nu$ \cite{GL 1985}, but is considerably 
more accurate. Since the chiral representation of the scalar form factor is
known to two loops 
\cite{Bijnens Colangelo Talavera}, the dependence of the correction
$\Delta_r$ on the quark masses is also known. In
addition to $\ell_1,\ldots\,,\ell_4$, the explicit expression involves
a further term, $r_{S2}$, from ${\cal L}_6$ \cite{BCE}. In the following,
we use this representation to eliminate the parameter $\ell_4$ in
favour of the scalar radius.

These numerical estimates of these low--energy constants are important when
analyzing the corrections to low--energy theorems. Let us look, e.g., at the
following two relations among the coefficients $c_1,\ldots\,,c_4$ which are
dictated by chiral symmetry. Consider the combinations
\bea \label{defC12}
\hspace*{-6em}C_1\all\equiv \all F_\pi^2\left\{c_2+4M_\pi^2(c_3-c_4)\right\}
\,,\hspace{0.5em}C_2\equiv \frac{F_\pi^2}{M_\pi^2}
\left\{-c_1+4M_\pi^4(c_3-c_4)\right\}\,.\eea
Chiral symmetry implies that, if the quark masses are turned off, both
$C_1$ and $C_2$ tend to 1. The contributions from $c_3$ and $c_4$ ensure that
the first order corrections only involve the symmetry breaking couplings
$\ell_3$ and $\ell_4$. Eliminating $\ell_4$ in favour of the scalar radius,
the low energy theorems take the form
\bea\label{letcr} 
C_1 \all=\all 1+
\frac{M_\pi^2}{3}\,\rs+\frac{23\,\xi}{420}+\xi^2 \Delta_1+O(\xi^3)\co\\
C_2\all=\all
1+\frac{M_\pi^2}{3}\,\rs+ \frac{\xi}{2}\left\{\lbar_3-
\frac{17}{21}\right\} +\xi^2 \Delta_2+O(\xi^3)\nonumber \fs\eea
At first nonleading order, $C_1$ is fully determined by 
the contribution from the scalar radius, while $C_2$ also contains a 
contribution from $\ell_3$.
Inserting the values $\rs=0.61 \,\mbox{fm}^2$ and $\lbar_3=2.9$ and ignoring 
the two-loop corrections $\Delta_1$, $\Delta_2$, we obtain
$C_1=1.103$, $C_2=1.117$. The value of $C_2$ differs little from $C_1$ --
as stated above, the estimate (\ref{l3barnum}) implies that the
contributions from $\ell_3$ are very small. 

\subsection{Phenomenological representation of the scattering amplitude}
If one neglects the contribution of the absorptive parts from $D$ waves
and higher, the $\pi \pi$ scattering amplitude can be written as a
combination of functions of a single variable that only have a right--hand
cut -- exactly like in the two--loop chiral representation. The imaginary
parts that appear inside the dispersive integrals will be evaluated from
the solutions of the Roy equations -- since these are based on
phenomenological information we will refer to this as the phenomenological
representation. We subtract the relevant dispersion 
integrals in the same manner as for the chiral representation:
\bea\label{dispWbar}\Wbar^I(s)\all=\all 
\frac{s^{4-\epsilon_I}}{\pi}\int_{4M_\pi^2}^{\infty}
ds'\;\frac{\mbox{Im}\,t^I(s')}
{s^{\prime\,4-\epsilon_I}(s-4M_\pi^2)^{\epsilon_I}(s'-s )}\fs\eea
Since all other contributions can be replaced by a polynomial, the
phenomenological amplitude takes the form
\bea\label{phenrep} A(s,t,u)\all=\all 16\pi a_0^2+
\frac{4\pi }{3M_\pi^2}\,(2a_0^0-5a_0^2)\,s+
\Pbar(s,t,u) \\\all\all +32\pi\left\{\mbox{$\frac{1}{3}$}
\Wbar^0(s)+\mbox{$\frac{3}{2}$}(s-u)\Wbar^1(t)
+\mbox{$\frac{3}{2}$}(s-t)\Wbar^1(u)\right.\no
\all\all \left.\hspace{2.3em}+\mbox{$\frac{1}{2}$}\Wbar^2(t)+
\mbox{$\frac{1}{2}$}\Wbar^2(u)
-\mbox{$\frac{1}{3}$} \Wbar^2(s) \right\}+O(p^8)\,.\nonumber\eea
We have explicitly displayed the contributions from the subtraction 
constants $a_0^0$ and $a_0^2$.
The term $\Pbar(s,t,u)$ is a crossing symmetric polynomial
\bea \Pbar(s,t,u)\all=\all \pbar_1+\pbar_2\,s+\pbar_3\,s^2+\pbar_4\,(t-u)^2+
\pbar_5\,s^3+\pbar_6\,s(t-u)^2\fs\eea
Its coefficients can be expressed in terms of integrals over the imaginary
parts of the partial waves. Explicit expressions can be found in
Ref.~\cite{CGL2}. In the following, the essential point is that the 
coefficients $\pbar_1,\ldots\,,\,\pbar_6$ can be determined
phenomenologically. 

\subsection{Matching the two representations}
In their common domain of validity, the two representations of the
scattering amplitude specified above agree, provided the parameters
occurring therein are properly matched:  
\bea \label{mc}C(s,t,u)=16\pi a_0^2+
\frac{4\pi }{3M_\pi^2}\,(2a_0^0-5a_0^2)\,s+\Pbar(s,t,u)+O(p^8)\fs \eea
Since the main uncertainties in the coefficients of the polynomial
$\Pbar(s,t,u)$ arise from their sensitivity to the scattering lengths
$a^0_0$, $a^2_0$, the above relations essentially determine the
coefficients $c_1,\ldots,\,c_6$ in terms of these two observables. The same
then also holds for the quantities $C_1$, $C_2$ defined in
eq.~(\ref{defC12}).  The corresponding low energy theorems for $a_0^0$ and
$a_0^2$ are of the form 
\bea \label{aC} a^0_0\all =\all
\frac{7M_\pi^2C_0}{32\pi F_\pi^2} +M_\pi^4\,\alpha_0
+O(m^4)\,,\hspace{0.5em} a^2_0=- \frac{M_\pi^2 C_2}{16\pi F_\pi^2}
+M_\pi^4\,\alpha_2 +O(m^4)\,, \eea 
with $C_0\equiv\frac{1}{7}(12\, C_1-5
\,C_2)$. The terms $\alpha_0$, $\alpha_2$ stand for integrals over the
imaginary parts of the partial waves that can be worked out from the
available experimental information. 
Formula (\ref{aC}) clearly shows how the matching works: chiral symmetry
provides accurate information on the unphysical quantities $C_0$ and $C_2$
(\ref{letcr}): the relation between these quantities and the scattering
lengths involves the terms $\alpha_0$ and $\alpha_2$, which are best
calculated from the explicit numerical solution of the Roy equations
discussed in the previous section. 

\subsection{Results for $a_0^0$ and $a_0^2$ and other threshold parameters}
Inserting the one-loop prediction for $C_1$, $C_2$ 
in the relations (\ref{aC}) and solving for
$a_0^0,a_0^2$, we obtain the following first order results:
\bea\label{one loop} a_0^0 = 0.2195\,,\hspace{0.5em}a^2_0 = -0.0446\,,
\hspace{0.5em}
2a_0^2-5a^2_0=0.662\,.\eea

The two-loop corrections $\Delta_1$ and $\Delta_2$ involve the coupling
constants $\ell_1,\,\ell_2,\,\ell_3$, the scalar radius, as well as the
terms $r_1,\ldots\,,r_4$, $r_{S2}$ from ${\cal L}_6$. The size of the
contributions from the latter may be estimated with the resonance model
described in Refs.~\cite{BCEGS,Bijnens Colangelo Talavera}.  The constants
$\ell_1$, $\ell_2$ can then be determined numerically with the
phenomenological values of $c_3$ and $c_4$. The resulting two-loop
corrections for the scattering lengths are very small.  The numerical
result is sensitive to the value of the scale $\mu$ at which the
renormalized coupling constants $r^r_n(\mu)$ are assumed to be saturated by
the resonance contributions. In the following, we use the resonance model
at the scale $\mu=M_\rho$ and take the range $500\mev \leq \mu \leq 1 \gev$
as an estimate for the uncertainties to be attached to the two-loop
corrections.

A careful evaluation of the uncertainties coming from the phenomenological
input leads to
\bea\label{l3rdependence} a_0^0\all=\all 0.220\pm 0.001
 -0.0017\,\Delta \ell_3+ 0.027\, \Delta_{r^2}\,,\\
a^2_0\all=\all -0.0444 \pm 0.0003 
-0.0004\,\Delta \ell_3-0.004\, \Delta_{r^2}\,,
\nonumber\eea
with $\lbar_3=2.9+\Delta \ell_3$, $\rs=0.61\,\mbox{fm}^2(1+\Delta_{r^2})$. 
Inserting the estimates (\ref{l3barnum}), (\ref{rsnum}), we arrive at our 
final result:
\bea\label{final result} \begin{array}{ll}
a_0^0= 0.220\pm 0.005\,,\all a^2_0=-0.0444\pm
0.0010\,,\\
2a_0^0-5a_0^2= 0.663\pm 0.006\,,\hspace{2em}\all a^0_0-a^2_0= 0.265\pm0.004\,.
\end{array}
\eea

\begin{table}[t]
\begin{center}
\begin{tabular}{|r|c|c|c|}
\hline
&Roy&CHPT
$O(p^6)$ Roy &CHPT $O(p^6)$ $K_{e4}$
\\ 
\hline
$a_0^0$&$ .220\pm .005$&$.215$&$.219\pm .005$\\
$-10 a_0^2$&$.444\pm .001$&$.445$&$.420\pm .010$\\
$10 b^0_0$&$2.76\pm 0.06$&$2.68$&$2.79\pm0.11$\\
$-10^2 b_0^2$&$8.03\pm 0.12$&$8.08$&$7.56\pm0.21$\\
$10^2 a_1^1$&$3.79 \pm 0.05$&$3.80$&$3.78\pm0.21$\\
$10^3 b_1^1$&$5.67\pm0.13$&$5.37$&$5.9\pm1.2$\\
\hline
$10^3 a^0_2$&$1.75\pm 0.03$&$1.76$&$2.2\pm0.4$\\
$10^4 a^2_2$&$1.70\pm 0.13$&$1.72$&$2.9\pm1$\\
\hline
\end{tabular}
\end{center}
\caption{\label{tab:threshold}
  Comparison of the numerical values of threshold parameters as
  obtained from the Roy solutions (after matching with CHPT), or CHPT. In
  the latter case the low--energy contants can be fixed either by the
  matching with Roy equations (third column), or by a phenomenological
  analysis of $K_{e4}$ form factors at two loops~\protect\cite{ABT_Ke4}
  (fourth column).}
\end{table}
Having fixed the subtraction constants of the Roy equations by matching to
the chiral representation we can now use the Roy solutions to evaluate the
$\pi \pi$ scattering amplitude at any energy. As an example we 
display in table \ref{tab:threshold} the results for other threshold
parameters. 
For comparison we also show the numerical values obtained from a direct
evaluation of the chiral two--loop amplitude. In this case we have to
specify how we fix the low--energy constants: since the matching procedure
does provide values for the latter, this is the most obvious choice. On the
other hand, in Ref.~\cite{ABT_Ke4} a two--loop analysis of the $K_{e4}$
form factors has lead to an alternative determination of the low--energy
constants. In table~\ref{tab:threshold} we give both sets of values for the
threshold parameters. The comparison shows clearly that two--loop chiral
perturbation theory works very well in describing both $\pi \pi$
scattering and $K_{e4}$ decays. On the other hand, the smallest error bars
for the threshold parameters are obtained by combining the chiral
representation and Roy equations. 

\section{Conclusions}
We have reviewed the current status of the theory of $\pi \pi$
scattering. The combination of numerical solutions of Roy equations and
two--loop chiral perturbation theory has lead to a remarkably precise
description of this reaction at low energy. As an example we have discussed
the numerical values of the two $S$--wave scattering lengths, which are
given with an uncertainty of a few percent, and of other threshold
parameters, with somewhat larger error bars.

The new experiments at Brookhaven \cite{e865} and at DA$\Phi$NE
\cite{kloe} on $K_{e4}$ decays will yield more precise information in the
very near future. On the basis of the preliminary data of
the E865 collaboration we have estimated in Ref.~\cite{CGL} the present
allowed range for $a_0^0$ to be between 0.20 and 0.25, in perfect agreement
with our prediction (\ref{final result}). Moreover, the pionic atom experiment
under way at CERN \cite{dirac} will allow a direct measurement of
$|a_0^0-a_0^2|$. These experimental tests will tell us whether we have
reached a full, accurate understanding of the low--energy dynamics of pion
interactions.

\section*{Acknowledgements} It is a pleasure to thank the organizers for
their excellent work, and B. Ananthanarayan, J. Gasser, H. Leutwyler and
G. Wanders for a most enjoyable collaboration.

\end{document}